\documentclass[%
reprint,
 amsmath,amssymb,
 aps,
]{revtex4-1}

\usepackage{graphicx}% Include figure files
\usepackage{dcolumn}% Align table columns on decimal point
\usepackage{bm}% bold math
\begin{document}

%\preprint{APS/123-QED}

\title{A random walker's view of networks whose growth it shapes}% Force line breaks with \\
%\thanks{A footnote to the article title}%

\author{Robert J.\@ H.\@ Ross}
\email{robert\_ross@hms.harvard.edu}
 %\altaffiliation[Also at ]{Physics Department, XYZ University.}%Lines break automatically or can be forced with \\
\author{Charlotte Strandkvist}%
 \email{charlotte\_strandkvist@hms.harvard.edu}
\author{Walter Fontana}%
 \email{walter\_fontana@hms.harvard.edu}
\affiliation{%
 Department of Systems Biology\\
 Harvard Medical School\\
 200 Longwood Avenue, Boston MA 02115
% This line break forced with \textbackslash\textbackslash
}%

%\date{\today}% It is always \today, today,
             %  but any date may be explicitly specified

\begin{abstract}
\noindent We study a simple model in which the growth of a network is determined by the location of one or more random walkers. Depending on walker speed, the model generates a spectrum of structures situated between well-known limiting cases. We demonstrate that the average degree observed by a walker is related to the global variance. Modulating the extent to which the location of node attachment is determined by the walker as opposed to random selection is akin to scaling the speed of the walker and generates new limiting behavior. The model raises questions about energetic and computational resource requirements in a physical instantiation.
%\begin{description}
%\item[PACS numbers]May be entered using the \verb+\pacs{#1}+ command.
%\end{description}
\end{abstract}

%\pacs{Valid PACS appear here}% PACS, the Physics and Astronomy
                             % Classification Scheme.
%\keywords{Suggested keywords}%Use showkeys class option if keyword
                              %display desired
\maketitle

%\tableofcontents

A large body of literature is devoted to analyzing systems that can be modeled as growing networks \cite{Porter2016,Hoffmann2013,Holme2012,Perra2012}.
In many cases network growth is coupled to a process situated on the network. Systems of this kind include the developing brain whose action potentials help shape neuronal architecture \cite{Bear}, social networks whose evolution is driven by interactions between the very individuals that constitute these networks, technological innovation which depends on current technologies within reach, and the internet whose structure is, among other things, determined by its usage \citep{Newman2010book}.

We study a simple network growth mechanism that is driven by a local process situated on the network. Our focus is to compare and relate the global view of the network with the local view of the network as accessible to a process situated on it. Many of the most influential network growth algorithms operate from a purely global perspective, that is, the entire network, or a statistic associated with it, is utilized in determining a growth event. For instance, the Barab\'asi-Albert growth algorithm requires knowledge of the global degree distribution. Similarly, exponential networks, and other more realistic models of network growth sample the location of a growth event from the entire network \cite{Dorogovtsev2000}. These issues have been previously noted \cite{Saramaki2004,Evans2005,Cannings2013,bloem2018random}, but it remains unclear when global growth strategies can be implemented by local processes subject to realistic constraints. Comparing local and global views will sharpen this question. To this end, we extend previous studies by using an exceedingly simplified local growth model based on a random walker. We focus on the expected degree of the node at which the walker is situated when a growth event occurs. This simple approach allows us to obtain conclusions regarding the degree distributions that can be generated by this model, and characterize canonical differences between global growth algorithms and algorithms enacted by local processes situated on the network.

We denote the number of nodes in the network at time $t$ with $V(t)\in\mathbb{N}$ and the number of edges with $E(t)\in\mathbb{N}$. $V(t)=N(t)+N_0$, where $N(t)$ is the number of growth events that occurred up to time $t$ and $N_0$ the number of nodes in the seed network. Each node $i$ is uniquely and permanently labelled at its creation by the count of growth events that have occurred up to and including its creation. Thus, $i \in \{ 1, 2, ..., V(t)\}$, with the last node labelled $V(t)$. The degree of node $i$ is denoted by $k_{i}$. 

In ``walker-induced network growth" (WING) a random walker situated on the network moves with a rate per time unit, $r_W$. In WING time is evolved continuously, in accordance with the Gillespie algorithm \cite{Gillespie_orig}, such that random walker movement and network growth events are modelled as exponentially distributed `reaction events' in a Markov chain. We employ a stochastic network growth mechanism in which the addition of a new (non-occupied) node occurs with rate $r_{N}$ per unit time. Network growth is therefore linear. At each growth event, a single node forms an edge with the node upon which the random walker is located. This means that just after addition to the network, new nodes are of degree $k = 1$. There is no limit to the number of nodes from which the network can be composed. We typically start the process with a fully connected network of $N_0$ nodes and a randomly chosen position of the walker (but nothing hinges on this choice, as shown in the Supplemental Material \cite{SM}).

\begin{figure}[h]
\includegraphics{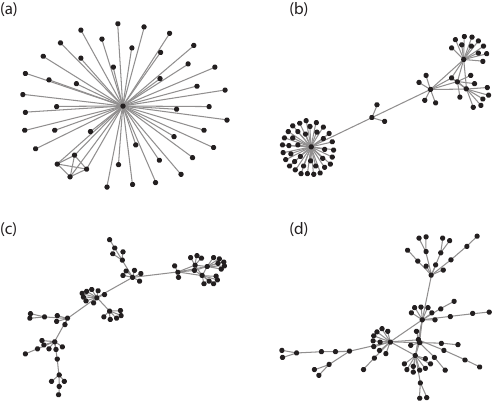}
\caption{\label{fig:networks} Networks created by WING. The motility rate, $r_W$, of the only walker differs across panels, while the network growth rate is constant $r_N=1$. For the sake of less congestion, simulations are stopped after $t=50$ time units. The seed network is a fully connected set of $N_0=5$ nodes, still recognizable as the only clique. Panel (a): $r_W = 0$; (b): $r_W = 0.1$; (c): $r_W = 1$; (d): $r_W = 100$.}
\end{figure}

The model can be extended in any number of ways and we shall consider two in particular. In one extension, the network hosts $m\le N_0$ walkers. When $m>1$, the walkers exclude each other in the sense that if a walker attempts to move to an already occupied node, the movement is aborted. The new node is connected to all $m$ distinct locations of the walkers and has therefore degree $m$. Although we mainly focus on WING obtained with a single walker, $m=1$, we touch occasionally on results obtained with multiple walkers. Networks generated with a single walker are necessarily trees (barring the initial seed network), whereas networks shaped by multiple walkers contain cycles. The other extension introduces a parameter $\alpha$, which modulates the coupling between walker and growth. With probability $\alpha$, growth occurs at the location of the walker as just described, and with probability $1-\alpha$ the new node is linked to a random location in the network.

\begin{figure}
\includegraphics{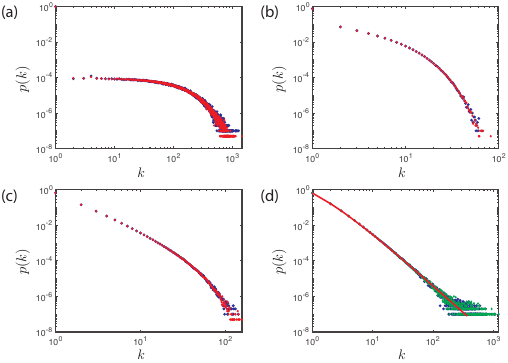}
\caption{\label{fig:degrees} Degree distributions generated by WING. The panels are for different $r_W$ values. For all panels $r_N=1$, $m=1$, and averages are taken over $1000$ replicates. Panel (a): $r_W=0.01$; (b): $r_W = 1$; (c): $r_W = 10$; (d): $r_W=100$. In panels (a)--(c), simulations were run to $t_{1} = 100,000$ (blue circles) and to $t_{2} = 200,000$ (red diamonds), indicating stationarity. In panel (d), representing the large-$r$ case, we compare the WING model (blue circles) with the BA procedure ($r_W\sim\infty$, green diamonds) at $t = 100,000$. The red line is $p_{BA}(k)=4/[k(k+1)(k+2)]$ as per \cite{Dorogovtsev2000,Krapivsky2000}.}
\end{figure}

Given an arbitrary initial network and in the absence of network growth ($r_N=0$), it is well-known that, in the long time limit, the probability $p_i$ of finding a random walker at node $i$ is proportional to its degree $k_i$, $p_i=k_i/\sum_j k_j=k_i/(2E)$, with $E$ the number of edges in the network. Barab\'asi and Albert (BA) considered a growing network \cite{Barabasi1999}, where growth occurs by repeatedly attaching a new node of degree $1$ to a node $i$ chosen according to $p_i(g)=k_i/[2E(g)]$ with $g$ indicating the growth step. This process, known as preferential attachment, was proposed as a mechanism for generating scale-free networks characterized by a power-law degree distribution $p(k)\sim k^{-\gamma}$. It was shown in \cite{Dorogovtsev2000,Krapivsky2000} that the BA procedure results in $p_{BA}(k)=4/[k(k+1)(k+2)]$. Preferential attachment hinges on a global view of the network, as it utilizes knowledge of $p_i(g)$ at every step $g$. This prompted Saram\"aki and others \cite{Saramaki2004,Evans2005} to propose a model in which random walkers sample the local connectivity of the growing network and serve as attachment points for the addition of a new node. The WING model shares the spirit of this approach, but is simpler and designed to study (i) how coupling an autonomous network growth to one or more random walkers on the network can yield networks of different kinds and (ii) how a network so-constructed might be described from the walker's point of view.

\begin{figure}[h]
\includegraphics[scale=0.8]{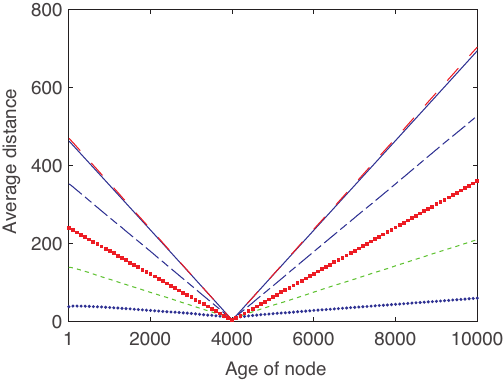}
\caption{\label{fig:distance} Distance between nodes as a function of age. Panel (a): The average distance between a node of age $4000$ and all other nodes in a network generated with WING is shown as a function of their age difference for different values of walker motility. $r_W = 10$ (blue disks), $r_W = 5$ (green dashed), $r_W = 2$ (blue dot-dashed), $r_W = 1$ (blue solid), $r_W = 0.5$ (red dashed) and $r_W = 0.1$ (red squares). $N=10,000$.}
\end{figure}

As expected, the network structures formed by the WING model depend on the ratio $r=r_W/r_N$, Figure \ref{fig:networks}. When $r_W=0$, the walker does not move from its initial position on the network and each new node therefore links to the same node, generating a ``star" structure, Figure \ref{fig:networks}a.  In the limit $r=r_W/r_N \rightarrow \infty$ the probability of finding the random walker at node $i$ becomes $p_i(t)=k_i/[2E(t)]$, which is the BA case, Figure \ref{fig:networks}d. This behavior is reflected in the degree distributions generated as $r$ sweeps across its range, as shown in Figure \ref{fig:degrees}. The main observations are: (i) At small $r$, the degree distribution is closer to an exponential, $p(k)\sim\exp(\beta k)$, reflecting the Poisson nature of attachment events, and approaches the power law of the BA case, $p_{BA}(k)$, as $r$ grows large; (ii) Stationarity, i.e.\@ a degree distribution independent of network size, is attained relatively quickly; (iii) Figure \ref{fig:distance} points to an interesting property of WING: a linear dependence between the average graph distance between two nodes $i$ and $j$ and their age difference, independent of network size. Distance is defined as usual in terms of the number of edges separating two nodes, while the age of a node $i$ is its label $N(i)$, a linear function of continuous time, and the age difference between two nodes $i$ and $j$ is $\vert N(j)-N(i)\vert$. The distance between nodes is maximized between $r_W = 0.5$ and $r_W = 1$.  This `big-world' property of the network is not observed for either the BA model or growing exponential models \cite{Dorogovtsev2000}.

In the Supplemental Material \cite{SM} we address the scenario when the network is grown by two walkers, $m=2$, and demonstrate that these results are robust to the shape of the seed network. We also address WING with exponential growth in the Supplemental Material \cite{SM}.

Because the walker shapes the growth of the network on which it moves, it is of interest to describe the network from the viewpoint of the walker. One way of doing so is to compute the average degree of the node the walker is situated at when a growth event occurs, as distinct from the average degree of a node given the network as a whole. We denote the walker's point of view with $\langle k_W\rangle$ and the global point of view with $\langle k\rangle$. We assume that WING attains a stationary degree distribution in the $N(t)\to\infty$ limit. This assumption is justified in the Supplemental Material \cite{SM}.

\begin{figure}[h!]
\includegraphics[scale=0.8]{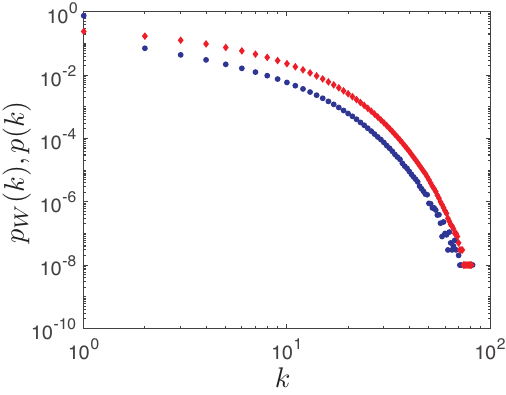}
\caption{\label{fig:localdegree} The local degree distribution. $p(k)$ (blue disks) is compared with $p_W(k)$ (red diamonds) after $N = 10,000$ growth events. Convergence of $p_W(k)$ to a stationary distribution is fast (Supplemental Material \cite{SM}). $r_W=1$, $r_N=1$.}
\end{figure}

Let $p_W(k)$ be the probability that the walker is situated at a node of degree $k$ when a growth event occurs. To obtain $p_W(k)$ numerically we record the trace (sequence) $\tau_r$ of degrees seen by the walker in a simulation $r$, collect the frequency with which the degree is $k$ at event $g$ across replicate traces $\tau_r$ ($r$ indexing the replicate) each comprising $N$ growth events, and average over $g$. Specifically, denoting the degree the walker observes at event $g$ of trace $\tau_r$ by $\tau_{r,g}$, we have
\begin{align}
    p_W(k,g) &= \dfrac{1}{R}\sum_{r=1}^R\delta(k-\tau_{r,g}) \nonumber\\
    p_W(k) &= \dfrac{1}{V}\sum_{g=1}^V p_W(k,g)
\end{align}
where $\delta(x)=1$ if $x=0$ and $\delta(x)=0$ otherwise. The global degree distribution $p(k)$ is computed likewise, but instead of observing a single node at growth event $g$, we observe all nodes in the network; thus, $p(k,g)$ is the probability of degree $k$ in a network at growth step $g$. Figure \ref{fig:localdegree} depicts $p_W(k)$ and $p(k)$ for $r=1$. Other values of $r$ generate similar plots (not shown). The walker sees a higher degree, except for degree $1$, which is the degree of newly created nodes.

We can express $\langle k_W\rangle$ as a function of global moments. Let $n(k,g)$ denote the average number of nodes of degree $k$ in a network when the $g$th growth event occurs. If $p(k)$ is stationary, $n(k,g)=g\,p(k)$. The change in $n(1,g)$ across a growth event is given by
\begin{eqnarray}\label{eq:grow1}
    n(1,g+1)-n(1,g)&=&(g+1)p(1)-g p(1)\nonumber\\
    &=&p(1)=1-p_W(1),
\end{eqnarray}
where the last equation follows because a growth event always adds one new node of degree $1$ and one node of degree $1$ is lost only if the walker is located at a degree-$1$ node, which happens with probability $p_W(1)$. Likewise, one node of degree $2$ is gained only if the walker is located at a degree-$1$ node and lost only if the walker is located at degree-$2$ node, yielding a net average change of $p(2)=p_W(1)-p_W(2)$. In general, we have the balance equations:
\begin{equation}\label{eq:grow2}
    p(k)=p_W(k-1)-p_W(k), \;\;k>1.
\end{equation}
From (\ref{eq:grow1}) and (\ref{eq:grow2}) we calculate the expected global degree:
\begin{equation}
    \langle k\rangle=\sum_{k=1}^{\infty}k\,p(k)=2.
    \label{eq:Nmoment1}
\end{equation}
In a similar fashion we obtain $\langle k^2\rangle=2 \langle k_W\rangle + 2$ and thus:
\begin{equation}
    \langle k_W\rangle=\dfrac{1}{2}\left(\langle k^2\rangle-2\right). \label{eq:Wmoment1}
\end{equation}
Interestingly, we found a relation similar to (\ref{eq:Wmoment1}) in \cite{Nimwegen1999}, although for non-growing networks. To summarize, from a global perspective, the average node degree of a network under WING is always $2$, regardless of the walker's rate of motion. In contrast, the walker observes an average node degree that reflects the global second moment of the network, which depends on $r_W$. For the second moment from the walker's perspective we obtain in a similar fashion:
\begin{equation}
    \langle k_W^2\rangle=\dfrac{1}{6}\left(2\langle k^3\rangle-3\langle k^2\rangle + 2\right). \label{eq:Wmoment2}
\end{equation}
In the Supplemental Material \cite{SM} we generalize these results to $m$ self-excluding walkers:
\begin{eqnarray}
    \langle k\rangle &=& 2m \nonumber \\
    \langle k_{W} \rangle &=& \dfrac{1}{2}\dfrac{\langle k^{2}\rangle - m(m+1)}{m} \nonumber \\
    \langle k_{W}^{2}\rangle &=& \dfrac{1}{3}\dfrac{\langle k^{3}\rangle - m^{3} - m - 3m\langle k_{W} \rangle}{m}, \label{eq:multi2}
\end{eqnarray}
where we used $\langle k_{W} \rangle$ in (\ref{eq:multi2}) for brevity.

\begin{figure}[h]
\includegraphics[scale=0.8]{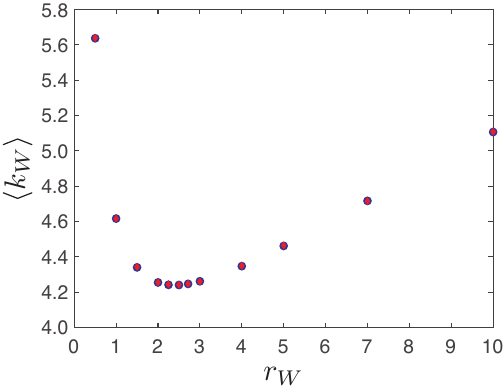}
\caption{\label{fig:meankW} The view from the walker. 
%The figure depicts the first moment of the degree distribution observed by the growth-inducing random walker as a function of its movement rate. 
The red diamonds show $\langle k_W\rangle$ obtained directly by simulation and the blue disks show $\langle k_W\rangle$ obtained from (\ref{eq:Wmoment1}) with $\langle k^2\rangle$ obtained from simulation. This should be contrasted with the expected node degree from the global perspective, (\ref{eq:Nmoment1}), which is independent of $r_W$.  The second moment, $\langle k^{2}_{W} \rangle$, is shown in the Supplemental Material \cite{SM}. $r_N=1$, $N=10,000$, and $R=100,000$ replicates.}
\end{figure}

Figure \ref{fig:meankW} shows $\langle k_W\rangle$ for $m=1$ as a function of walker motility. 

Together with Figure \ref{fig:meankW}, equations (\ref{eq:Wmoment1}) and (\ref{eq:Wmoment2}) permit a few observations. (i) $\langle k_W\rangle$ is finite for $0<r_W<\infty$. (ii) $\langle k_W\rangle$ diverges in both limits $r_W\rightarrow\infty$ and $r_W\rightarrow 0$. As $r_W\rightarrow\infty$, WING approaches the BA procedure, which yields power law $p_{BA}(k)$ with divergent $\langle k^2\rangle$. In the limit $r_W\rightarrow 0$, the walker is pinned and generates a star network with a single node of divergent degree. (iii) Since $\langle k_W\rangle$ is a convex function of $r_W$, oscillations could be generated by a mechanism in which the walker's motility is a function of the observed mean degree. Moreover, $\langle k_W\rangle$ is minimized when $\langle k^2\rangle$ is minimized. A more sophisticated walker could achieve this minimization by employing a gradient descent with respect to its own motility. (iv) A given $\langle k_W\rangle$ can be attained with a pair of distinct $r_W$ values, which, by (\ref{eq:Wmoment1}), yield global $p(k)$-distributions with the same $\langle k^2\rangle$ (and variance, since $\langle k\rangle=2$ always). Yet, these $r_W$ values do not yield the same $\langle k_W^2\rangle$, Figure S3 in \citep{SM}, and hence the $p(k)$-distributions must differ in the third moment $\langle k^3\rangle$ by virtue of (\ref{eq:Wmoment2}). 

In the Supplemental Material \cite{SM} we argue that for $0 \le r_W < \infty$ the WING model cannot generate a Barab\'asi-Albert network with stationary degree distribution $p_{BA}(k)$, as already suggested by Figure \ref{fig:degrees}. In the Supplemental Material \cite{SM}, we show that WING admits a stationary degree distribution for any $r_W$.

We next refine the WING model by adding a parameter $0 \le\alpha\le 1$ that tunes the influence of the walker on network growth: A growth event links the new node with probability $\alpha$ to the location of the walker, and with probability $1-\alpha$ to a node chosen uniformly at random from the network, which includes the location of the walker. The case described so far corresponds to $\alpha = 1$. Modulating $\alpha$ in this manner allows us to explore network growth algorithms that contain both global and local aspects in their computation.

\begin{figure}[h]
\includegraphics[scale=0.8]{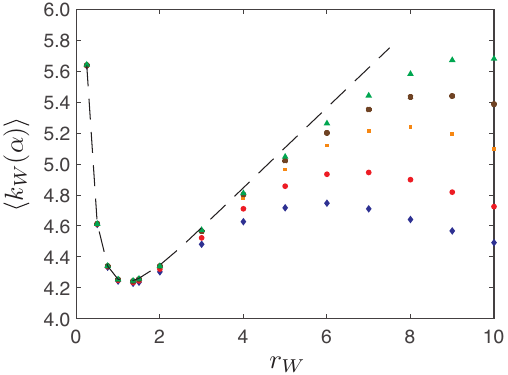}
\caption{\label{fig:alpha} Tunable walker influence. The graph depicts $\langle k_{W}(\alpha)\rangle$ as a function of walker motility $r_W$ for different time points (numbers of growth events or, equivalently, network sizes) $N$, with $r_N = 1$ and $\alpha = 0.5$. Data points are averaged over $10,000$ replicates. The simulation is used to acquire $\langle k^{2}(\alpha) \rangle$, while $\langle k_{W}(\alpha)\rangle$ is computed via (\ref{eq:alpha1st}). We obtain the same graphs using both simulation data and (\ref{eq:alpha1st}), Supplemental Material \cite{SM}. The local moment equations thus appear to apply even at shorter time scales when the global degree distribution is not stationary. Blue diamonds: $N=10,000$, red disks: $N=20,000$, orange squares: $N=50,000$, brown asterisks: $N=100,000$, green triangles: $N=200,000$. The coupling parameter $\alpha=0.5$ has here half the value it has in Figure \ref{fig:meankW}. The dashed line is the graph shown in Figure \ref{fig:meankW} but plotted against an abscissa scaled by a factor of $1/\alpha=2$.}
\end{figure}

Following the same reasoning of the last section, we compute the first and second moments of the degree distribution $p_W(k,\alpha)$ observed by the walker as a function of $\alpha$:
\begin{eqnarray}
   \langle k_{W}(\alpha)\rangle &=& \dfrac{1}{2\alpha}\left(\langle k^{2}(\alpha) \rangle + 4\alpha - 6\right)
\label{eq:alpha1st}\\
   \langle k_{W}^2(\alpha)\rangle &=& \dfrac{1}{3\alpha}\left(\langle k^{3}(\alpha)\rangle + 3\langle k^{2}(\alpha)\rangle\left(\alpha - \dfrac{3}{2}\right) + 1\right).
\label{eq:alpha2nd}
\end{eqnarray}
These results are generalized in the Supplemental Material \cite{SM} to $m>1$. In Figure \ref{fig:alpha} we compare (\ref{eq:alpha1st}) for different network sizes $V$. The following observations stand out for the modified WING model. 

(i) Figure \ref{fig:alpha} indicates that convergence for $\alpha<1$ at high walker motility is slow compared to $\alpha=1$ or low walker motility. At low walker motility, the networks tend to have exponential degree distributions (Figure \ref{fig:networks}) and lowering $\alpha$ only contributes to this structure. Given that (\ref{eq:alpha1st}) was derived on the assumption of stationarity, it is remarkable how well the local first moment, $\langle k_{W}(\alpha)\rangle$, tracks the global second moment, $\langle k^{2}(\alpha) \rangle$, far from stationarity (Supplemental Material \cite{SM}), suggesting a quasi-stationary global degree distribution $p(k)$. The slowly evolving global statistical structure of the network must be replicated locally for the walker to sense it and follow (\ref{eq:alpha1st}).

(ii) The dotted line in Figure \ref{fig:alpha} is Figure \ref{fig:meankW} with the abscissa scaled by a factor of $2$: the average degree seen by the walker at speed $r_W$ when growth always occurs at the location of the walker is the same as that seen by a walker with twice that speed when growth occurs half the time at the location of the walker and half the time at a random location. This suggests that
\begin{equation}
    \langle k_{W}(\alpha_1=1,r_W)\rangle=\langle k_{W}(\alpha_2,r_W/\alpha_2)\rangle,
    \label{eq:motilityscaling}
\end{equation}
where the dependency on $r_W$ is made explicit and $r_N=1$. As network size $V(t)$ tends to $\infty$ in the limit, the probability that a growth event occurs at the location of the walker tends to $\alpha$, since it becomes increasingly unlikely that any random growth events, occurring with probability $1-\alpha$, hit the walker's location by chance. The walker can therefore be viewed as having an effective motility or $\alpha$-horizon $\hat{r}_W=r_W/\alpha$. We can rephrase (\ref{eq:motilityscaling}) as asserting that the fraction $1-\alpha$ of growth events do not affect, in the large $V(t)$ limit, the world that the walker sees within its ``$\alpha$-horizon".

(iii) It follows that
\begin{equation}
    % \lim_{\alpha\to 0}\langle k_{W}(\alpha)\rangle \ne \langle k_{W}(0)\rangle.
        \lim_{\alpha\to 0}\langle k_{W}(\alpha)\rangle \rightarrow \infty.
\end{equation}
The effective motility of the walker, equation (\ref{eq:motilityscaling}), diverges in the limit $\alpha\to 0$, yielding the BA procedure with a divergent second global moment and therefore a divergent $\langle k_W(\alpha)\rangle$ as in the case of $\alpha=1$ and $r_W\to\infty$.
% To determine $\langle k_W(\alpha=0)\rangle$, consider that when we set $\alpha=0$, we have in the limit $V\to\infty$:
% \begin{equation}
%     p_W(k)=\dfrac{k\,p(k)}{\sum\limits_{j=1}^{\infty}j\,p(j)},
% \end{equation}
% where we have dropped the explicit mention of $\alpha$ to declutter. This is because the probability of the walker being at a particular node in a random network is proportional to the degree $k$ of that node and $p(k)$ is the probability that such a degree is realized. When $\alpha=0$, the network has an exponential degree distribution \cite{Dorogovtsev2013}: $p(k) = 2^{-k}$. Hence, $\sum_{k=1}^{\infty}k\,p(k)=\sum_{k=1}^{\infty}k2^{-k}=\langle k\rangle=2$ and 
% \begin{equation}
%     \langle k_W\rangle=\dfrac{1}{\langle k\rangle}\sum_{k=1}^{\infty}k^2 2^{-k}=
%     \dfrac{\langle k^2\rangle}{\langle k\rangle}=\dfrac{6}{2}=3.
% \end{equation}

(iv) The behavior as $r_W\to\infty$ and the behavior at $r_W=\infty$ differ when $\alpha<1$ compared with $\alpha=1$. From our treatment of the $\alpha=1$ case and observation (iii) above, $\langle k_W\rangle$ diverges for any $\alpha$ as $r_W\to\infty$. However, if we set $r_W=\infty$, the random walk is theoretically treated as always in equilibrium on the network, i.e.\@ there is no concept of an $\alpha$-horizon for the walker at any $\alpha<1$, and $\langle k_W\rangle$ remains finite due to the fraction $1-\alpha$ of random growth events. For $\alpha=1$, $\langle k_W\rangle$ diverges.
\\
\\
We have presented a simple network growth mechanism in which random walkers on a network control where the network grows, and thus determine its structure.  Many real-world networks exhibit structures that are determined, at least in part, by processes situated on them \cite{Newman2010book}.  We provided a description of WING by taking the perspective of the walker, expressing the expected degree of the node the walker is situated at when a growth event occurs, $\langle k_{W} \rangle$, and demonstrating that when $0 < r_W < \infty$, $\langle k_W \rangle$ remains finite.  We then extended the model by adding a parameter $\alpha$ controlling the coupling between network growth and walker location, and showed that this gives rise to an effective motility that causes new behavior as $\alpha\to 0$ and $r_W\to\infty$.

It is important to address whether WING could effectively generate the network constructed by the BA model. That is, given the requirement for the position of the random walker to equilibrate over an increasingly large network, is this energetically feasible?  Even though at finite values of $r_W$ the degree distribution associated with the BA model is approximately achieved, as demonstrated by Figure \ref{fig:degrees},  it seems network growth algorithms that are formulated from a global perspective of the network may require unrealistic behaviours if they are to be enacted by local processes situated on the network. The transition between an algorithm based on global criteria being efficiently implementable by local processes at small but not at large network sizes, might also be of interest for identifying whether real-world networks are being built using global knowledge or local processes, or why a network may appear to grow differently once it has exceeded a certain size. The modulation of $\alpha$ allowed us to examine network growth algorithms that utilize both local processes and global knowledge of the network. In this scenario it is apparent that both the degree distribution and the observations of the random walker become a function of the network size (Figure \ref{fig:alpha}). This is only captured, however, when the finite motility of the random walker is taken into account. A different interpretation would be arrived at if the walker's position was treated at equilibrium on the network, which could be seen to correspond to a global perspective such as that found in the BA model.  This observation serves to demonstrate that equilibrium assumptions must be made with caution when applied to local processes known to guide a network's growth.  

In view of Figure \ref{fig:meankW}, it would be interesting to determine the value of $r_W$ at which $\mathrm{d} \langle k_{W} \rangle/\mathrm{d}r_W = 0$. This the point at which the difference between $\langle k_W \rangle$ and $\langle k \rangle$ is minimized, and appears numerically close to Euler's number. More generally, since $r_W$ determines the structure of the network the walker observes, there may be interesting further work into simple mechanisms by which the walker can control what it observes, without resorting to global knowledge.  Alternatively, we may ask in what ways an external force can control the observations of a random walker situated on a growing network.  For instance, if $\alpha$ were to be made a function of network size, would a naive random walker be able to discriminate between a change in $\alpha$ and a change in its own motility without recourse to an intrinsic notion of time?  Finally, we have only studied how the position of an unbiased random walker evolves, which we did for mathematical ease. However, many other processes could be studied, such as proliferating random walkers of different types interacting on the same growing network \cite{Khain2007,Ross2016b,Ross2017b}.

\noindent RJHR would like to thank Ioana Cristescu and Pavel Krapivsky for helpful discussions.

%\bibliography{references}

%merlin.mbs apsrev4-1.bst 2010-07-25 4.21a (PWD, AO, DPC) hacked
%Control: key (0)
%Control: author (8) initials jnrlst
%Control: editor formatted (1) identically to author
%Control: production of article title (-1) disabled
%Control: page (0) single
%Control: year (1) truncated
%Control: production of eprint (0) enabled
%

\pagebreak
\widetext
\clearpage

\begin{center}
\textbf{\large A random walker's view of networks whose growth it shapes\\Supplemental Material}
\end{center}
%%%%%%%%%% Merge with supplemental materials %%%%%%%%%%
%%%%%%%%%% Prefix a "S" to all equations, figures, tables and reset the counter %%%%%%%%%%
\setcounter{section}{0}
\setcounter{equation}{0}
\setcounter{figure}{0}
\setcounter{table}{0}
\setcounter{page}{1}
\makeatletter
\renewcommand{\theequation}{S\arabic{equation}}
\renewcommand{\thefigure}{S\arabic{figure}}
\renewcommand{\bibnumfmt}[1]{[S#1]}
\renewcommand{\citenumfont}[1]{S#1}
\renewcommand{\thesection}{\Alph{section}}

\section{Random initial networks}
\label{supp:initial}

\begin{figure}[!h]
\begin{center}
\includegraphics[scale=0.65]{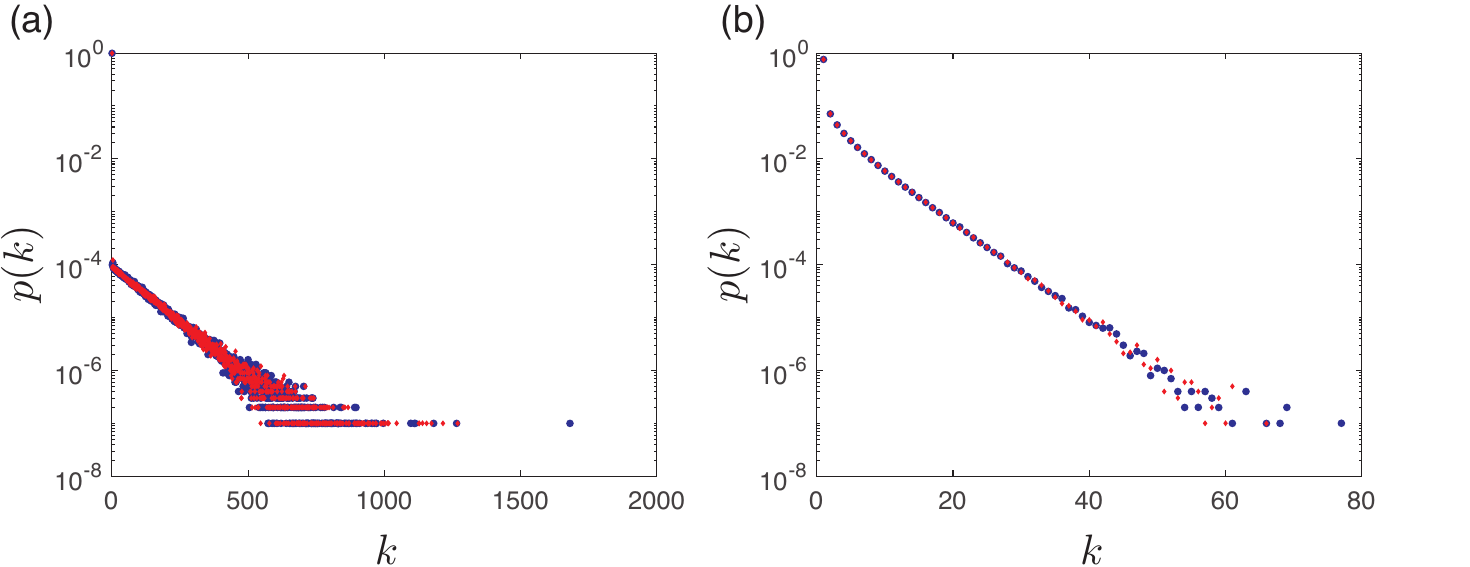}
\caption[Degree distribution from random initial networks]{Degree distributions from random initial networks (blue disks) compared to fully connected initial networks (red diamonds). Panel (a): $r_W = 0.01$. Panel (b): $r_W = 1$. In all panels $m=1$, $r_N=1$, $N = 100,000$; averages from $1,000$ replicates.}
\label{suppfig:initial}
\end{center}
\end{figure}

In Figure \ref{suppfig:initial} it is shown that initializing WING dynamics with random networks has little effect on the degree distribution compared to initializing with fully connected networks.

\section{Convergence of simulations}
\label{supp:convergence}

\begin{figure}[h!]
\begin{center}
\includegraphics[scale=0.55]{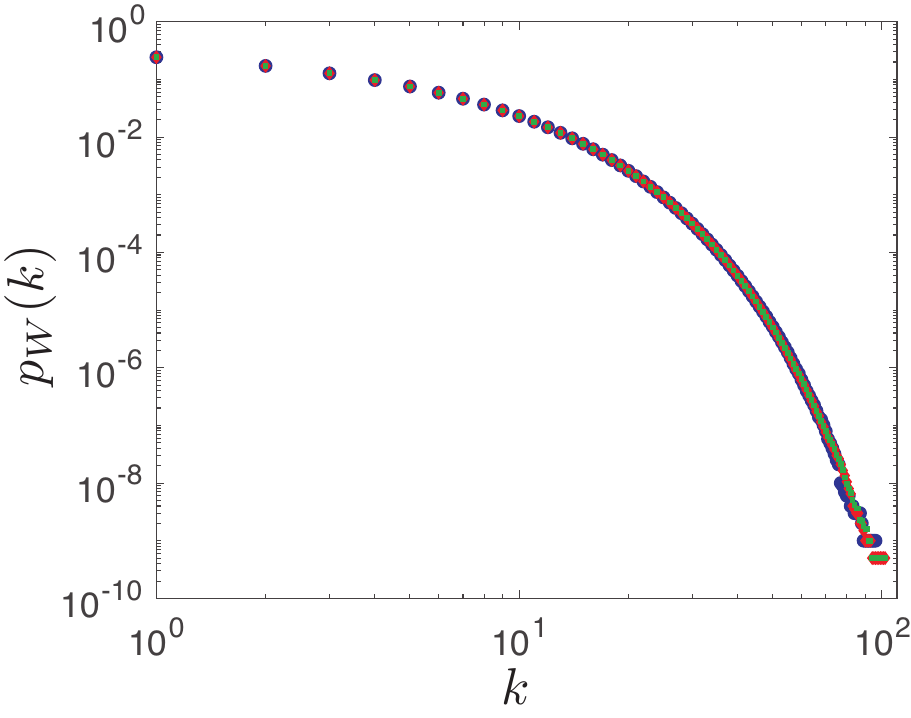}
\caption[Convergence to the local degree distribution]{Convergence to the local degree distribution. $p_W(k)$ is depicted for different growth extents: $N=10,000$ (blue disks), $N=20,000$ (red diamonds), and $N=100,000$ (green squares). $r_W=1$, $r_N=1$, $m=1$, $\alpha=1$.}
\label{suppfig:localdegreeB}
\end{center}
\end{figure}

Figure \ref{suppfig:localdegreeB} indicates that convergence on a stationary local degree distribution for the basic WING model ($m=1,\alpha=1$) is fast. This Figure accompanies Figure 4 in the main text.

\newpage
\section{Second moment of the local degree distribution}
\label{supp:secondlocal}

This Figure accompanies Figure 5 in the main text.

\begin{figure}[ht]
\begin{center}
\includegraphics[scale=0.55]{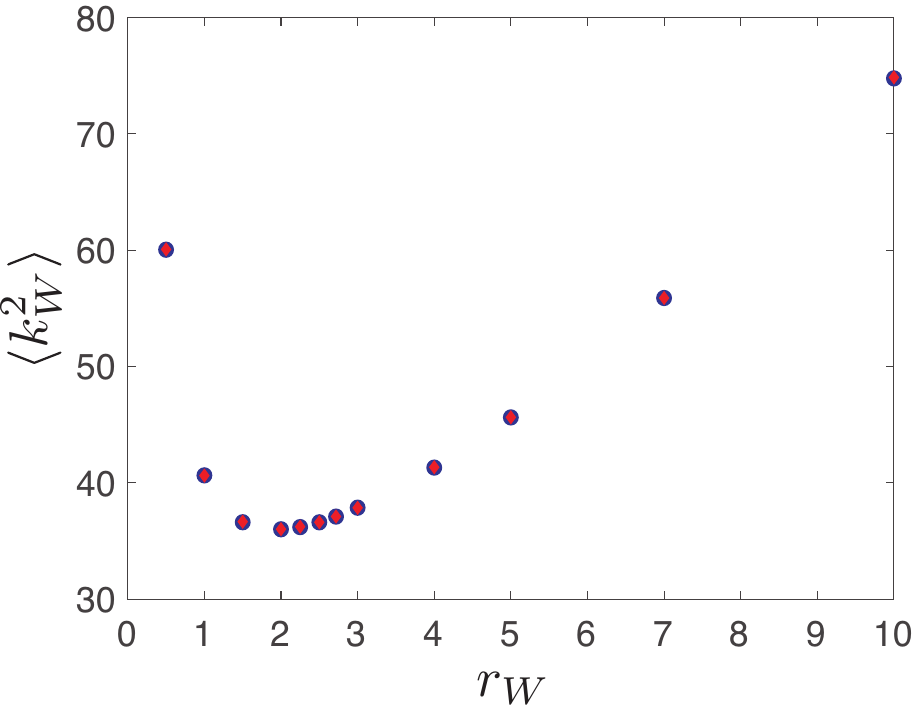}
\caption[Second walker moment]{Second walker moment. The figure depicts the second moment of the degree distribution observed by the random walker as a function of its movement rate. The red diamonds show $\langle k_W^2\rangle$ obtained directly by simulation and the blue disks show $\langle k_W^2\rangle$ obtained from equation (6) in the main text with $\langle k^2\rangle$ and $\langle k^3\rangle$ obtained from simulation. $r_N=1$, $N=10,000$, and $R=100,000$ replicates.}
\label{suppfig:meankWB}
\end{center}
\end{figure}

\section{Applicability of local moment equations at short time scales}
\label{supp:secondlocal2}

This Figure accompanies Figure 6 in the main text.

\begin{figure}[!h]
\begin{center}
\includegraphics[scale=0.5]{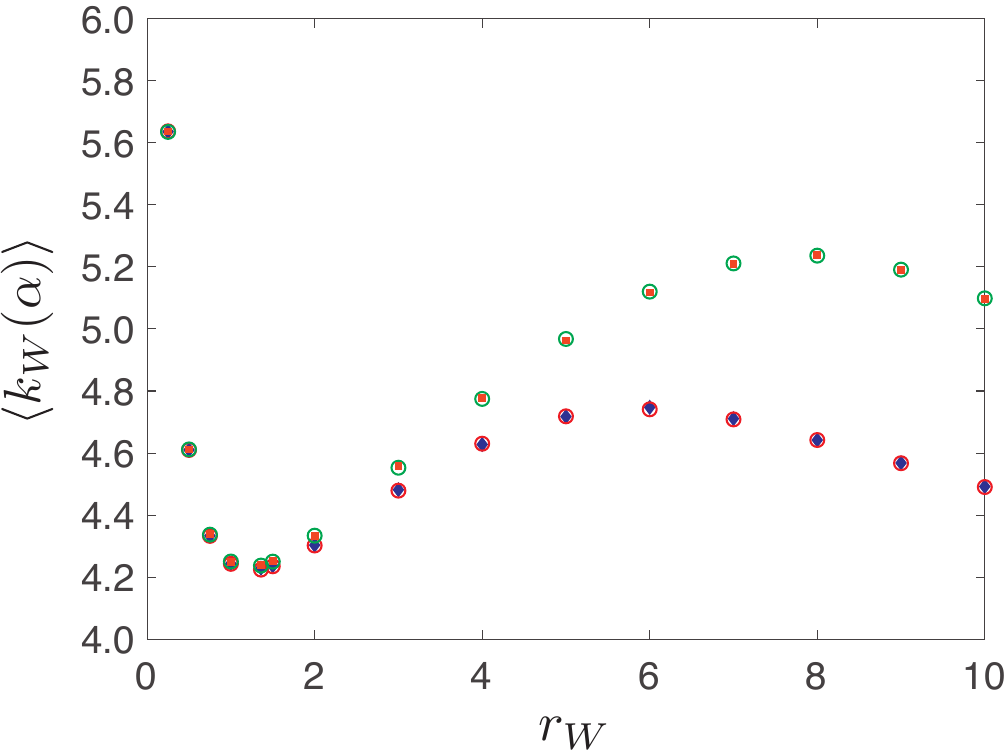}
\caption[Tunable walker influence]{Tunable walker influence. The graph demonstrates the applicability of the local moment equations even at shorter time scales when the global degree distribution is not stationary. $\langle k_{W}(\alpha)\rangle$ is plotted as a function of walker motility $r_W$ for different time points (numbers of growth events or, equivalently, network sizes) $N$ using $r_N = 1$ and $\alpha = 0.5$. The blue diamonds ($N=10,000$) and orange squares ($N=50,000$) show $\langle k_{W}(\alpha)\rangle$ computed via (8) in the main text, based on the same simulation data as in panel (a) of Figure 6 of the main text. The red and green circles show $\langle k_{W}(\alpha)\rangle$ directly obtained from simulation data. Data points are averaged over $10,000$ replicates.}
\label{suppfig:alphaB}
\end{center}
\end{figure}

\section{The BA degree distribution cannot be the stationary distribution for any finite motility in WING}
\label{supp:noBA}

Let $w=r_W/(r_W+r_N)$ and $g=r_W/(r_W+r_N)$ denote the probabilities that the next event is either a step by the walker or the growth of the network, respectively. Furthermore, let $p_{N(t)}$ denote the probability that, when a growth event occurs, the random walker is located at the last node added to the network. If the stationary distribution were $p_{BA}(k)$, $p_{N(t)}$ would tend to zero in the limit $t\to\infty$ (i.e.\@ $N(t)\to\infty$), because all nodes in the network must be visited proportional to their in-degree. 
%We know from equation (5) in the main text that the average degree observed by the walker, $\langle k_W \rangle$, behaves like the second moment of the global degree distribution, $\langle k^{2} \rangle$, which diverges for $p_{BA}(k)$. 
Hence $p_{N(t)}$ must vanish. Yet, for $r_W<\infty$, $p_{N(t)}$ is bounded below:
\begin{align}
    p_{N(t)}>\sum_{k=1}^{\infty}p_W(k)\,g\,\dfrac{w}{k+1}.
\end{align}
where $p_W(k)$ is the stationary degree distribution from the walker's viewpoint. The right hand side represents the network-size independent probability of just one scenario for the walker to be positioned on the last node added: the walker is at a node of degree $k$, a growth event occurs, and the walker moves to the added node. Clearly, there are many more ways for the walker to reach that node before the next growth event occurs, especially when $w$ is large (hence $g$ small). However, a lower bound for $p_{N(t)}$ contradicts its vanishing implied by the assumption that $p_{BA}(k)$ is the global stationary degree distribution. Hence $p_{BA}(k)$ cannot be the stationary degree distribution for WING for any $r_W<\infty$.

\section{WING dynamics admits a stationary degree distribution}
\label{supp:stationary}

By a stationary degree distribution $p(k)$ we mean that
\begin{align}\label{eq:prob}
    \lim_{N(t) \rightarrow \infty}\dfrac{n(k,t)}{N(t)} \rightarrow p(k,t)=p(k) \text{ constant in $t$},
\end{align}
where $n(k,t)$ is the number of nodes of degree $k$ in a network of $N(t)$ nodes. We have $n(k,t)=N(t)\,p(k,t)$ and $d\,n(k,t)/dN(t)=p(k,t)+dp(k,t)/dN(t)=p(k,t)$ in the large-$N(t)$ limit since $p(k,t)$ does not depend on $N(t)$ as limit cycles or other complex behaviors do not arise in WING dynamics by construction. 

The same reasoning that led to equations (2) and (3) in the main text yields without stationarity assumption
\begin{align}
    \dfrac{dn(1,t)}{dN(t)}&=p(1,t)=1-p_W(1,t)\\
    \dfrac{dn(k,t)}{dN(t)}&=p(k,t)=p_W(k-1,t)-p_W(k,t),\,\,k>1.
\end{align}
As in the main text, these balance equations express that nodes of degree $k$ are lost by linking to the new node at a rate $p_W(k,t)$, the probability that the walker is at a node of degree $k$ just prior to a growth event. For $r_W>0$, $p_W(1,t)$ cannot go to zero or no nodes of degree $1$ would ever be lost, yielding the star network for which $p(1,t)=1$ for all $t$, which is attained only when $r_W=0$. Hence $p_W(1,t)$ is bounded below by some $c_1 wg$. 
%The bound would be $wg$ if every step of the walker prior to a growth event was to a node of degree $1$, but there are many more sequences of steps to reach a degree-$1$ node prior to growth. 
This entails $p_W(k,t)>c_k wg^k$, and by conservation $p_W(k,t)\ge p_W(k+1,t)$. Since by virtue of the lower bounds $p_W(k,t)\to\alpha_k$ as $N(t)\to\infty$, we have that $p(1,t)\to 1-\alpha_1$, $p(2,t)\to \alpha_1-\alpha_2$, and so on.

\section{Exponential network growth}
\label{supp:exp}

\begin{figure}[!h]
\begin{center}
\includegraphics[scale=0.55]{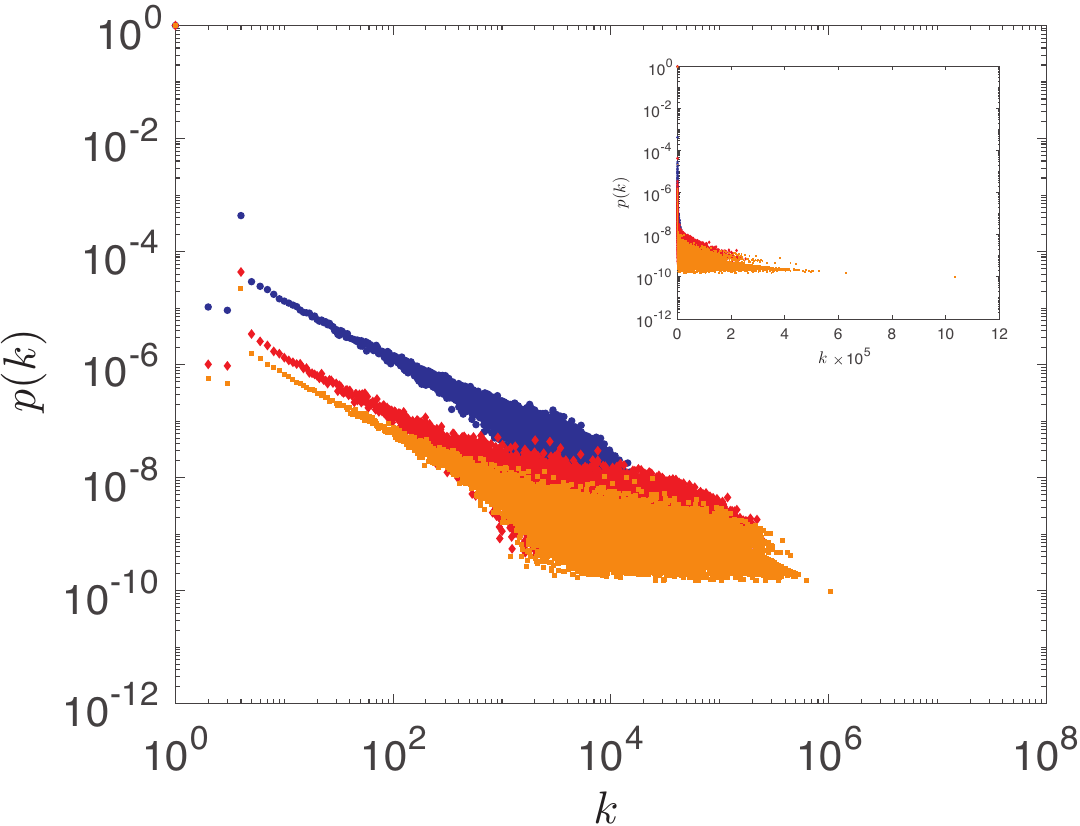}
\caption[Exponential network growth and constant walker motility]{Exponential network growth and constant walker motility. Degree distributions generated by exponential WING dynamics. Networks consist of (on average) $10,000$ nodes (blue disks), $100,000$ nodes (red diamonds), $200,000$ nodes (orange squares). Averages are over $1,000$ replicates with $r_W=1$, $r_N=1$, $m=1$. The inset shows the same data on a lin-log scale.}
\label{suppfig:exponential}
\end{center}
\end{figure}

In exponential WING dynamics, the network grows with a rate constant $r_N$ per node, which is to say an overall growth rate of $r_N N$. Figure \ref{suppfig:exponential} indicates that under these conditions $p(1)\to 1$, that is uniqueness and non-trivial stationarity of the degree distribution are lost.

\begin{figure}[!h]
\begin{center}
\includegraphics[scale=0.65]{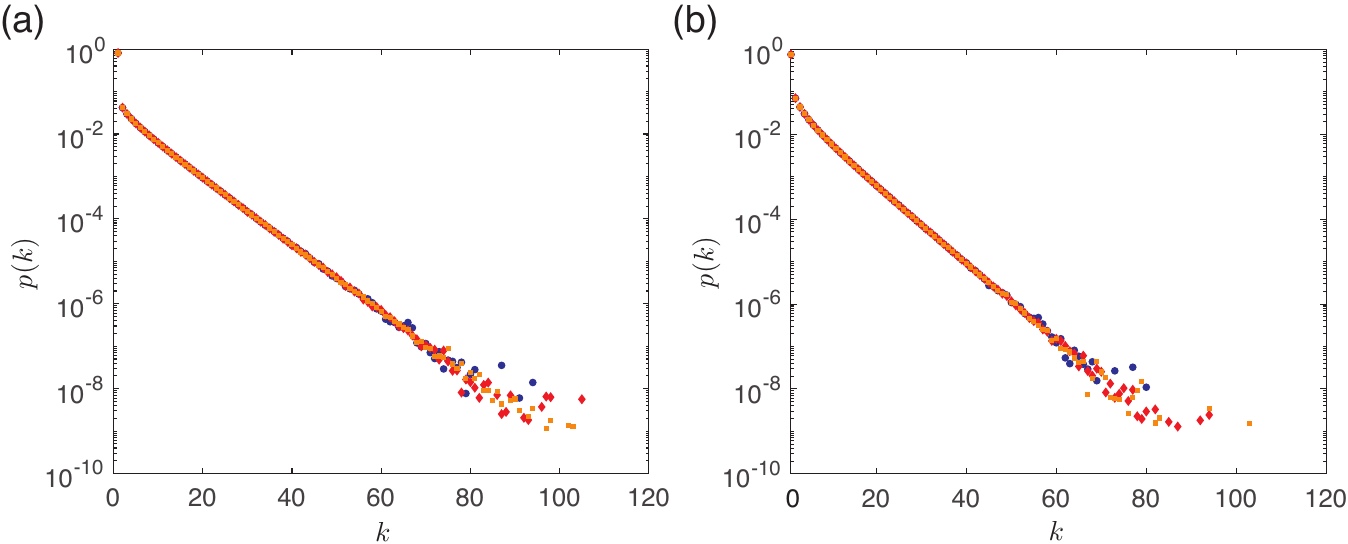}
\caption[Exponential network growth and proportional walker motility]{Exponential network growth and proportional walker motility. Degree distributions generated by exponential WING dynamics with walker motility proportional to network size. Networks consist of (on average) $10,000$ nodes (blue disks), $100,000$ nodes (red diamonds), $200,000$ nodes (orange squares). Averages are over $1,000$ replicates with $r_N=1$, $m=1$. Panel (a): $r_W=N+1$. Panel (b): $r_W=(N+1)/2$.}
\label{suppfig:exponentialR}
\end{center}
\end{figure}

When network growth is linear, the effective growth rate is constant, $\lim_{N \rightarrow \infty}r_W/r_N = c$, whereas it tends to zero when growth is exponential, $\lim_{N \rightarrow \infty}r_W/(r_N N) = 0$. This suggests that if the random walker motility were proportional to network size, $O(N)$, stationarity and uniqueness of the WING degree distribution should be restored.  Figure \ref{suppfig:exponentialR} shows that this is indeed the case.

\section{Multiple walkers}
\label{supp:multiwalk}

\begin{figure}[!h]
\begin{center}
\includegraphics[scale=0.55]{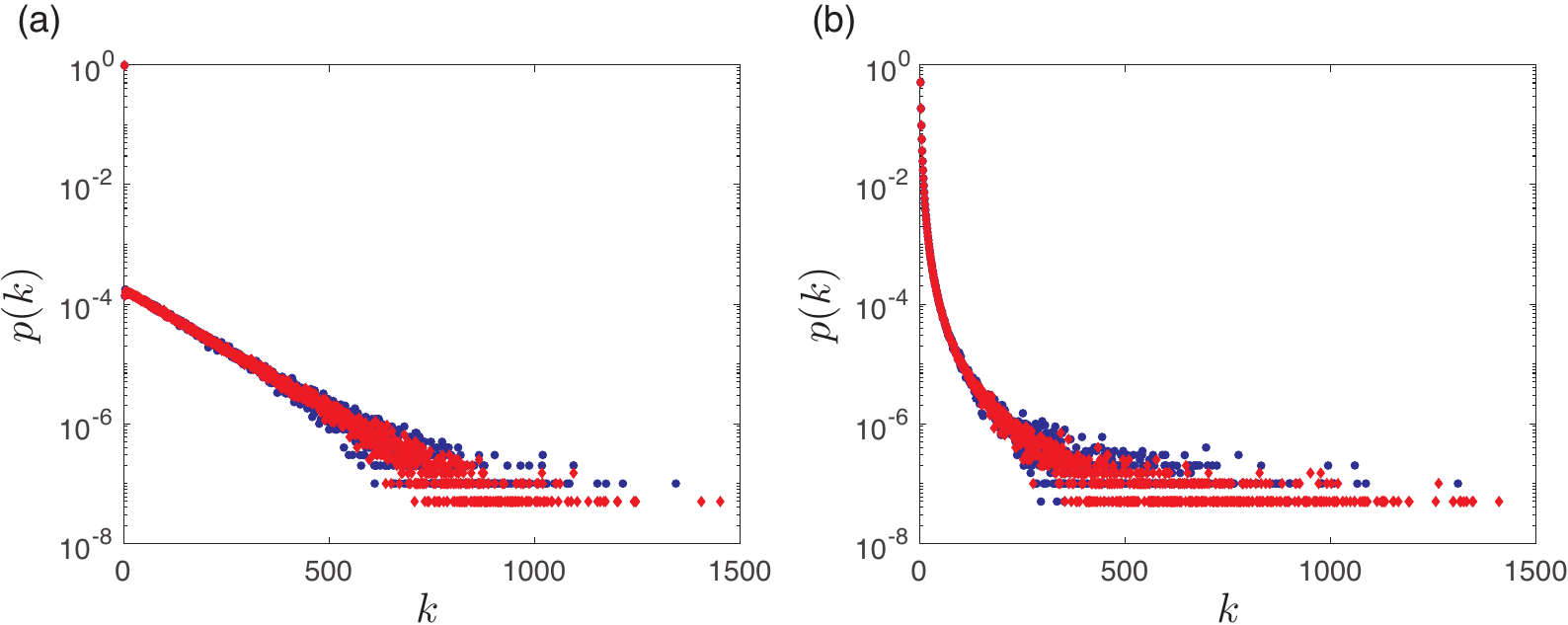}
\caption[Degree distribution from multiple walkers]{Degree distribution from multiple walkers. Examples of degree distributions generated by WING dynamics with $m=2$. Panel (a): $r_W = 0.01$. Panel (b): $r_W = 10$. In all panels  $r_N=1$, $N = 100,000$ (blue disks), $N = 200,000$ (red diamonds); averages from $1,000$ replicates.}
\label{suppfig:mdegrees}
\end{center}
\end{figure}

Figure \ref{suppfig:mdegrees} indicates that multiple walkers, $m>1$, generate stationary degree distributions as in the case of $m=1$.

We next solve the balance equations (2) and (3) in the main text for the case of $m$ walkers. The balance equation for the global probability of a node having degree $z$ depends on the joint probabilities of finding $1\le l\le m$ of $m$ walkers at nodes with degree $z$ and, similarly, at nodes with degree $z-1$. Let the (stationary) joint probability of finding the $m$ walkers at nodes with degrees $k_1,k_2,\ldots k_m$ be $\hat{p}(k_1,\ldots,k_m)$, where $\hat{p}(\cdot)$ is with probability $1-\alpha$ the global degree distribution $p(\cdot)$ in case the growth event occurs at a random node and with probability $\alpha$ the degree distribution $p_W(\cdot)$ from a walker's perspective (when growth occurs at the location of the walkers). There are $\binom{m}{l}$ ways of choosing $l$ walkers to be placed on nodes of degree $z$, each choice having probability 
\begin{align}
    \sum_{k_1\ne z}^{\infty}\cdots\sum_{k_{m-l}\ne z}^{\infty}\hat{p}(k_1,\ldots,k_{m-l},\underbrace{z,\ldots,z}_{\text{$l$ times}})
\end{align}
where we canonicalized $\hat{p}(\cdot)$ by assigning the highest subscripts to degrees fixed at $z$. Each of the combinations affects $l$ nodes of degree $z$ and the contribution (positive or negative depending on the specific balance equation considered) to the change in $n(z)$ across a growth event is therefore given by
\begin{align}\label{eq:mwalk1}
    \sum_{l=1}^m l \binom{m}{l}\sum_{k_1\ne z}^{\infty}\cdots\sum_{k_{m-l}\ne z}^{\infty}\hat{p}(k_1,\ldots,k_{m-l},\underbrace{z,\ldots,z}_{\text{$l$ times}})
\end{align}
This expression can be simplified considerably. To this end we split (\ref{eq:mwalk1}) into a contribution from $l=1$ and the rest:
\begin{align}
    m\sum_{k_1\ne z}^{\infty}\cdots\sum_{k_{m-1}\ne z}^{\infty}\hat{p}(k_1,\ldots,k_{m-l},z)+\sum_{l=2}^m l \binom{m}{l}\sum_{k_1\ne z}^{\infty}\cdots\sum_{k_{m-l}\ne z}^{\infty}\hat{p}(k_1,\ldots,k_{m-l},z,\ldots,z).\label{eq:mwalk2}
\end{align}
The second expression of sums in (\ref{eq:mwalk2}) provides exactly all terms excluded in the first expression. To see this, recast the first expression of (\ref{eq:mwalk2}) explicitly in terms of combinations $\hat{p}(k_1,\ldots,k_{m-1},z),$ $\hat{p}(k_1,\ldots,k_{m-2},z,k_{m}),\ldots, \hat{p}(z,\ldots,k_{m})$ with implied summation, excluding the value $z$, over all degree variables $k_i\ne z$. For $l=2$, the second expression contains the $\binom{m}{2}$ combinations of $\hat{p}(\cdot)$ with two $z$, such as $\hat{p}(k_1,z,k_3,\ldots,k_{m-2},z,k_{m})$.
This particular combination, for example, supplies all terms excluded in $\hat{p}(k_1,z,k_3,\ldots,k_{m})$ of the first expression. It also does so for all terms excluded in $\hat{p}(k_1,k_2,\ldots,k_{m-2},z,k_{m})$. In general, any of the $\binom{m}{l}$ combinations supplies the excluded $\hat{p}(\cdot)$ terms for all the $\binom{m}{l-1}$ combinations from which it differs in the position of exactly one $z$, of which there are $l$ instances. Hence, the sum (\ref{eq:mwalk2}) is but the marginal of the joint
\begin{align}
    &\,m\sum_{k_1=1}^{\infty}\cdots\sum_{k_{m-1}=1}^{\infty}\hat{p}(k_1,\ldots,k_{m-1},z)\\
    =&\,m\,\hat{p}(z).\label{eq:simple1}
\end{align}
With this simplification, the balance equations read
\begin{align}
    p(1) &= \delta(1-m)-m(1-\alpha)p(1)-m\alpha p_W(1)\\
    p(k) &= \delta(k-m)+m(1-\alpha)[p(k-1)-p(k)] + m\alpha [p_W(k-1)-p_W(k)],\quad k>1.
\end{align}
The $\delta(k-m)$ term accounts for the fact that a node of degree $m$ is always added to the network since the incoming node connects to all $m$ walkers.

We proceed similarly to the $m=1$ case in the main text to calculate
\begin{align}
    \langle k\rangle = 2m,
\end{align}
and
\begin{align}
    \langle k^2 \rangle &= m^{2} + m(1-\alpha)(2 \langle k \rangle + 1) + m\alpha(2 \langle k_W \rangle + 1)
\end{align}
from which we obtain
\begin{align}\label{eq:kWmulti}
    \langle k_{W} \rangle = \dfrac{1}{2}\dfrac{\langle k^{2}\rangle - m((m+1) + 2\langle k\rangle (1 - \alpha))}{m\alpha}.
\end{align}
Note that $\langle k_{W} \rangle$ is an average across all $m$ walkers. For $\alpha=1$ and $m=2$ equation (\ref{eq:kWmulti}) becomes
\begin{align}
    \langle k_{W} \rangle = \dfrac{1}{4}(\langle k^{2}\rangle - 6).
\end{align}

The second moment is given by 
\begin{align}\label{eq:k2Wmulti}
    \langle k_{W}^{2}\rangle = \dfrac{1}{3}\dfrac{\langle k^{3}\rangle - m^{3} - m(1-\alpha)(3\langle k^{2}\rangle + 3\langle k \rangle + 1) - m\alpha - 3m\alpha\langle k_{W} \rangle}{m\alpha},
\end{align}
where we used $\langle k_{W} \rangle$, equation (\ref{eq:kWmulti}), on the right hand side for brevity.

\begin{figure}[!h]
\begin{center}
\includegraphics[scale=0.65]{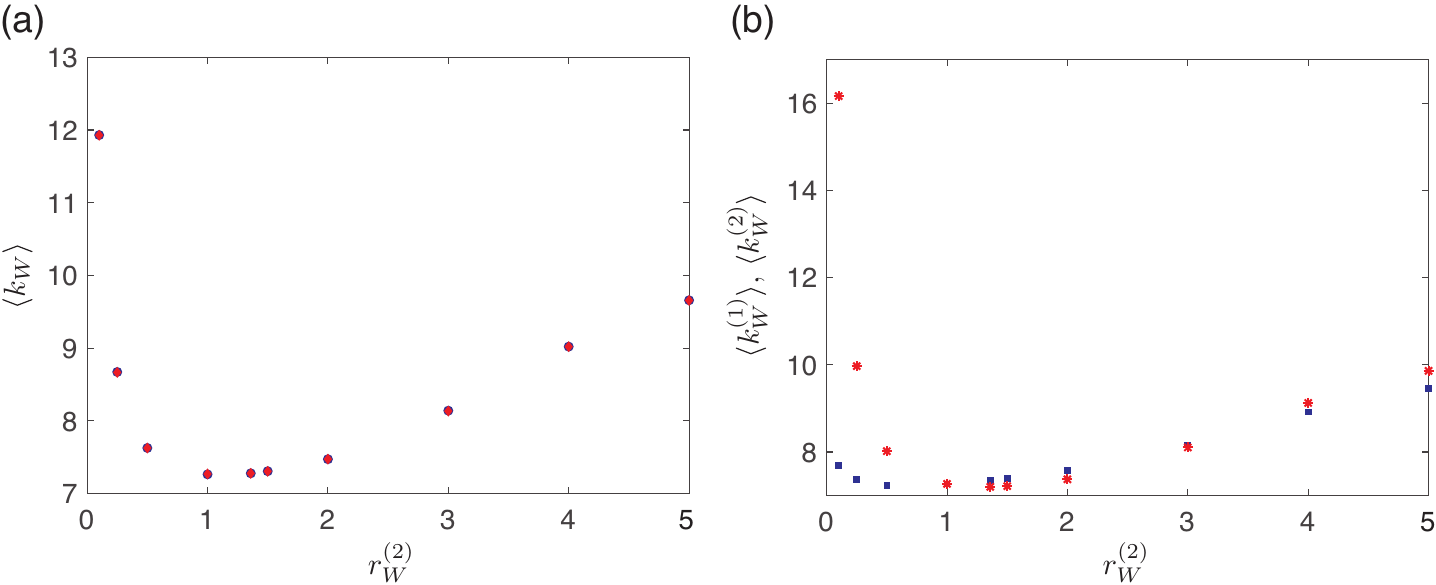}
\caption[Multiple walkers]{Multiple walkers. The figure depicts the first moment of the degree distribution observed by two random walkers, $1$ and $2$, as a function of the movement rate of walker $2$. Panel (a): The red diamonds show $\langle k_W\rangle$ averaged over both walkers obtained directly by simulation and the blue disks show $\langle k_W\rangle$ according to (\ref{eq:kWmulti}) with $m=2$ and $\alpha=1$, where $\langle k^{2}\rangle$ is obtained from simulation. Panel (b): The mean degree seen by each walker as obtained from simulation directly. In both panels, $r_N=1$, $N=2,000,000$, and $R=100$ replicates.}
\label{suppfig:mwalk}
\end{center}
\end{figure}

Figure \ref{suppfig:mwalk} shows that equation (\ref{eq:kWmulti}) agrees extremely well with simulations.
Figure \ref{suppfig:mwalk} and equations (\ref{eq:kWmulti}) and (\ref{eq:k2Wmulti}) permit the following observations. (i) The expected degrees of the nodes at which walker $1$ and $2$ are located when a growth event occurs, $\langle k_{W}^{(1)} \rangle$ and $\langle k_{W}^{(2)} \rangle$, respectively, are not the same as if each walker was on the network alone, or if two nodes of degree $k = 1$ were added to the network for each growth event. It is also apparent from Figure \ref{suppfig:mwalk} that $\langle k_{W}^{(1)} \rangle$ and $\langle k_{W}^{(2)} \rangle$ have both the same value twice. (ii) When $r_W^{(1)} = c$, where $c$ is a constant, and $r_W^{(2)} \rightarrow 0$, both $\langle k_{W}^{(1)} \rangle$ and $\langle k_{W}^{(2)} \rangle$ diverge in the limit.  However, if $r_W^{(1)} = c$ and $r_W^{(2)} = 0$, then only $\langle k_{W}^{(2)} \rangle$ will diverge and $\langle k_{W}^{(1)} \rangle$ remains finite. (iii) When $r_W^{(1)} = c$ and $r_W^{(2)} = \infty$ (theoretically), both $\langle k_{W}^{(1)} \rangle$ and $\langle k_{W}^{(2)} \rangle$ remain finite. This is also true in the limit when $r_W^{(1)} = c$ and $r_W^{(2)} \rightarrow \infty$. (iv) Following a growth event the random walkers will ``meet" at a constant rate in the limit $N \rightarrow \infty$. The scenario generating a lower bound for this is given by a growth event, followed by random walker $1$ moving to the new node, followed by random walker $2$ attempting to move to the new node but being excluded by walker $2$. 
A lower bound can be written as:
\begin{align}\label{eq:bump}
gw^{(1)}\left(\sum_{k=1}^{\infty}\frac{p^{1}_{W}(k)}{k+1}\right)w^{(2)}\left(\sum_{k=1}^{\infty}\frac{p^{2}_{W}(k)}{k+1}\right),
\end{align}
which is independent of $N$.

\end{document}